\documentclass[12pt,a4paper]{article}
\pdfoutput=1

\usepackage{epsfig}
\usepackage{amsmath}
\usepackage{amssymb}
\usepackage{verbatim}
\usepackage{bm}
\usepackage{dsfont}
\usepackage[footnotesize]{caption}
\usepackage{subfigure}

\begin{document}

\begin{flushright}
\tt{MPP-2010-79}
\end{flushright}

\vskip 0.05cm

\begin{center}
{\Large\bf Sneutrino Hybrid Inflation and \\[1mm] Nonthermal Leptogenesis}

\vskip 1cm

Stefan~Antusch, Jochen~P.~Baumann, Valerie~F.~Domcke and \\ Philipp~M.~Kostka\\[3mm]
{\it{
Max-Planck-Institut f\"ur Physik (Werner-Heisenberg-Institut),\\
F\"ohringer Ring 6,
80805 M\"unchen, Germany}
}
\end{center}

\vskip 0.75cm

\begin{abstract}
In sneutrino hybrid inflation the superpartner of one of the right-handed neutrinos involved in the seesaw mechanism plays the role of the inflaton field. It obtains its large mass after the ``waterfall'' phase transition which ends hybrid inflation. After this phase transition the oscillations of the sneutrino inflaton field may dominate the universe and efficiently produce the baryon asymmetry of the universe via nonthermal leptogenesis. We investigate the conditions under which inflation, with primordial perturbations in accordance with the latest WMAP results, as well as successful nonthermal leptogenesis can be realized simultaneously within the sneutrino hybrid inflation scenario. We point out which requirements successful inflation and leptogenesis impose on the seesaw parameters, i.e.\ on the Yukawa couplings and the mass of the right-handed (s)neutrino, and derive the predictions for the CMB observables in terms of the right-handed (s)neutrino mass and the other relevant model parameters. 
\end{abstract}

\newpage

\section{Introduction}

The paradigm of cosmic inflation~\cite{Guth:1980zm, Liddle:2000cg, Bailin:2004zd} (for recent reviews see e.g.~\cite{Baumann:2009ds, Mazumdar:2010sa}) has proven very successful in resolving the flatness and horizon problems of the early universe and in explaining the absence of relics from early phase transitions. However, the connection to particle physics is still unclear. One possibility to establish such a connection is provided by sneutrino hybrid inflation~\cite{Antusch:2004hd}, where the superpartner of one of the right-handed neutrinos involved in the seesaw mechanism~\cite{seesaw} plays the role of the inflaton field. In sneutrino hybrid inflation, a large vacuum energy density is present which drives inflation and the sneutrino direction in field space has an almost flat potential suitable for slow-roll inflation. The right-handed (s)neutrinos obtain their large masses after the ``waterfall'' phase transition which ends hybrid inflation. Inflation in this scenario is closely linked to the physics generating the small neutrino masses via the seesaw mechanism. 

Another attractive connection between the seesaw mechanism and early universe cosmology is the possibility of generating the observed baryon asymmetry of the universe via the out-of-equilibrium decays of the right-handed (s)neutrinos in leptogenesis~\cite{Fukugita:1986hr} (for recent reviews see~\cite{Chen:2007fv, Davidson:2008bu}). For calculating the produced baryon asymmetry, the knowledge of the phase of (p)reheating after inflation is in general mandatory, since it may lead to the nonthermal production of right-handed (s)neutrinos and since it determines the reheat temperature, which in turn governs the possibility of thermal (s)neutrino production. In most inflation models the nonthermal (s)neutrino production must arise from the decays of the inflaton field. In sneutrino hybrid inflation, on the other hand, the inflaton itself is a right-handed sneutrino, which means that this intermediate step is skipped and the sneutrino inflaton field after inflation may directly dominate the universe and, when it decays, most efficiently produce the baryon asymmetry and reheat the universe.

In previous works, leptogenesis after sneutrino inflation has been studied in the context of chaotic sneutrino inflation~\cite{Murayama:1992ua, Ellis:2003sq} which however requires a quite heavy sneutrino with a mass of about $10^{13}$ GeV and correspondingly very small Yukawa couplings in order to realize a low reheat temperature as suggested by gravitino and similar constraints in supersymmetric cosmology. Furthermore, chaotic sneutrino inflation with a quadratic potential for the inflaton gives rise to a comparatively large tensor-to-scalar ratio of $r \sim 0.16$.
On the other hand, sneutrino hybrid inflation~\cite{Antusch:2004hd}, as typical for hybrid-type inflation scenarios~\cite{Linde:1990gz, Linde:1991km, Copeland:1994vg, Linde:1997sj}, predicts a much smaller ratio $r \lesssim 0.01$ and is thus clearly distinguishable from chaotic sneutrino inflation by future observations (e.g.\ by the Planck satellite).  Recently, it has been shown that sneutrino hybrid inflation~\cite{Antusch:2004hd} belongs to a wider class of hybrid-like inflation models, dubbed ``tribrid inflation'' in~\cite{Antusch:2009vg}, which are very suitable for being embedded into supergravity (SUGRA) theories with the SUGRA $\eta$-problem solved by either a shift symmetry
\footnote{In the context of chaotic inflation, shift symmetry has been 
used e.g. in~\cite{Kawasaki:2000yn, Yamaguchi:2000vm, Kawasaki:2000ws}.}~\cite{Antusch:2009ef} or a Heisenberg symmetry~\cite{Antusch:2008pn} in the K\"ahler potential. While the sneutrino was a gauge singlet in~\cite{Antusch:2004hd}, it has been demonstrated in~\cite{Antusch:2010va} that it may be embedded into a Grand Unified Theory (GUT) representation, e.g.\ into a \textbf{16}-plet of SO(10), establishing a possible link between sneutrino hybrid inflation and left-right symmetric GUTs. Nonthermal leptogenesis after sneutrino hybrid inflation, on the other hand, was only briefly discussed in~\cite{Antusch:2004hd} for an example set of model parameters.

In this paper, we therefore investigate in detail the conditions under which inflation, with primordial perturbations in accordance with the latest WMAP results, as well as successful nonthermal leptogenesis can be realized simultaneously within the  sneutrino hybrid inflation scenario. We point out which requirements successful inflation and leptogenesis impose on the seesaw parameters, i.e.\ on the Yukawa couplings and the mass of the right-handed (s)neutrino, and derive the predictions for the CMB observables in terms of the right-handed (s)neutrino mass and the other relevant model parameters. Our results are meant as a guideline for the construction of explicit particle physics models incorporating sneutrino hybrid inflation and baryogenesis via nonthermal leptogenesis. 

The paper is organized as follows: In section~\ref{framework} we introduce the sneutrino hybrid inflation scenario in a simple setup. Section~\ref{inflation} is dedicated to the inflationary phase and the predictions for the CMB observables. In section~\ref{leptogenesis} we discusses the reheating of the universe after inflation and the production of the baryon asymmetry of the universe via nonthermal leptogenesis. We conclude in section~\ref{infl_lepto} by combining our results from inflation and leptogenesis to highlight the preferred ranges of the model parameters.

\section{Framework}\label{framework}

We will discuss sneutrino hybrid inflation and subsequent baryogenesis via nonthermal leptogenesis in an extension of the  minimal supersymmetric standard model (MSSM) with conserved R-parity, 
where three additional right-handed (s)neutrinos acquire large masses after the ``waterfall'' phase transition ending inflation. The superpotential defining our framework is given by
\begin{equation}\label{eq_W}
W = W_{\text{MSSM}} + (y_{\nu})_{ij} \,\hat{N}^i\, \hat{h}_{ a}\, \epsilon^{a b} \hat{L}^j_b  
+ \frac{\lambda_{ii}}{M_P}\,(\hat{N}^i)^2 \hat{H}^2 + \kappa\, \hat{S}\left(\hat{H}^2 - M^2\right) + \ldots\,,
\end{equation}
where the $\hat{N}^i$ (the index $i=1, 2, 3$ denotes the different generations) are gauge singlet superfields describing the heavy right-handed (s)neutrinos and where the reduced Planck scale is given by $M_P\simeq2.4\cdot 10^{18}\,\text{GeV}$. The canonically normalized imaginary parts\footnote{At this point we will assume  that inflation proceeds along the imaginary direction of the complex scalar sneutrino field. We will see later that this can be obtained by a shift symmetry in the K\"ahler potential, which protects this direction against the SUGRA $\eta$-problem as was demonstrated in~\cite{Antusch:2009ef}.} $N^i$ of the respective scalar components are inflaton candidates as will be described below. $\hat{L}$ and $\hat{h}$ are SU(2)$_L$-doublet superfields which contain the standard model leptons and up-type Higgs, respectively. The Yukawa coupling term of $\hat{N}$ with the Higgs and lepton doublet, i.e.\ the second term in Eq.~\eqref{eq_W}, allows to identify $\hat{N}$ with the right-handed neutrino superfield.

$\hat{H}$ and $\hat{S}$ are two additional gauge singlet superfields. Here the canonically normalized real part $H$  of the scalar component of $\hat{H}$ is the so-called ``waterfall'' field responsible for ending inflation. The F-term of $\hat{S}$, the so-called ``driving superfield'', provides the large vacuum energy density that drives inflation. The scalar component of $\hat{S}$ is fixed at zero during inflation by SUGRA corrections (cf.\ section \ref{sugra}) and does not affect the inflationary dynamics. Furthermore, we assume $\lambda_{ii}$ and $\kappa$ to be real coupling parameters for simplicity.

The form of the superpotential Eq.~\eqref{eq_W} is motivated as follows: The latter two terms generate the scalar potential suitable for inflation. In the false vacuum with large values of $N^i$ and $H$ stabilized at zero, the large vacuum energy $V_0=\kappa^2 M^4$ drives the quasi-exponential growth of the scale factor in inflation. Once the slow-rolling fields $N^i$ fall below a critical value,  the negative contribution to the squared mass of $H$ from the term $\kappa\, \hat{S}\,(\hat{H}^2 - M^2)$ starts dominating over the positive contribution from the terms $ \frac{\lambda_{ii}}{M_P}\,(\hat{N}^i)^2 \hat{H}^2$. Therefore, $H$ becomes tachyonic which triggers the ``waterfall'' ending inflation as $H$ acquires a non-zero vacuum expectation value (vev).

\begin{figure}
\center
\includegraphics[scale=0.9]{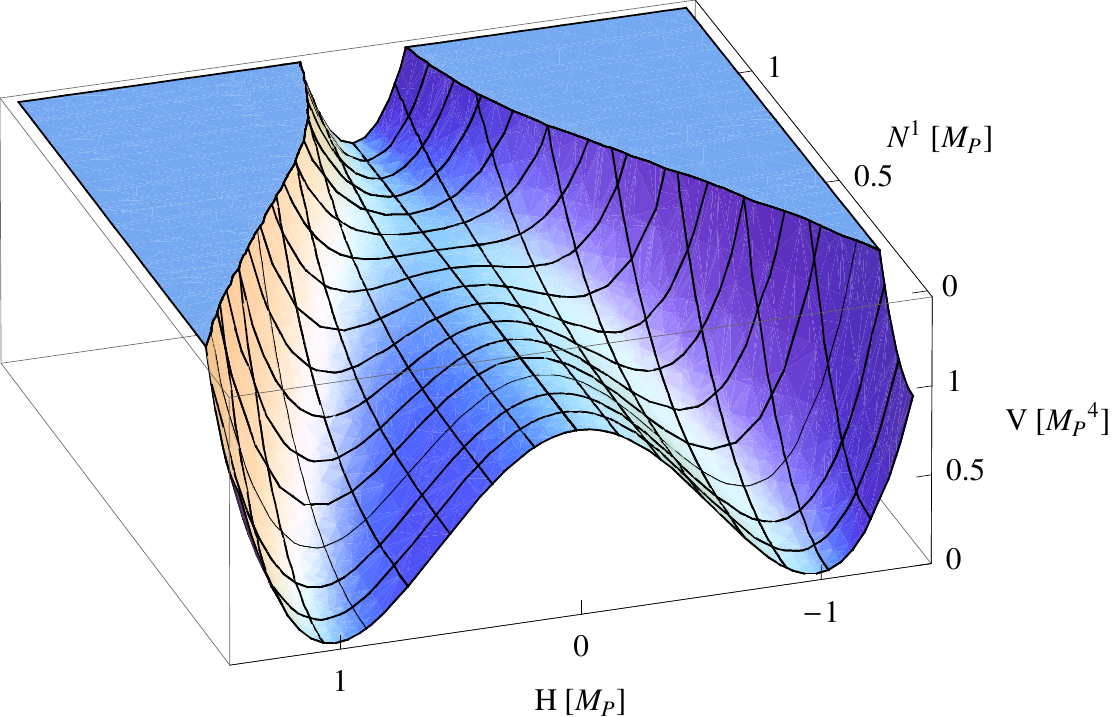}
\caption{\label{hybridpotential} Typical F-term scalar potential resulting from the model of Eq.~\eqref{eq_W} for the scalar components of $\hat{S},\, \hat{L},\, \hat{h}$ set to zero. For the plot, we have used example parameters $\kappa=\lambda=1$ and $M=M_P$.}
\end{figure}

After inflation, close to the global minimum of the potential where $N^i \approx 0$ and $H \approx \sqrt{2} \,M$ and where the large vacuum energy contribution vanishes, the fields $N^i$ and $H$ perform damped oscillations accounting for a matter dominated universe.  The field which decays last and finally dominates the universe is generically the right-handed sneutrino with the smallest mass and smallest Yukawa couplings, since it only decays via the second term in Eq.~\eqref{eq_W} proportional to $(y_{\nu})_{ij}$. This decay reheats the universe which thus enters its radiation dominated epoch. For illustration, we have plotted a typical scalar potential resulting from the scenario described above in Fig.~\ref{hybridpotential}. For further details see e.g.~Ref.~\cite{Antusch:2009ef}.

The first three terms in Eq.~\eqref{eq_W} describe the MSSM with masses for the additional right-handed neutrinos generated after inflation. In particular the term $\frac{\lambda_{ii}}{M_P}(\hat{N}^i)^2 \hat{H}^2$ generates mass terms for the heavy (s)neutrinos as $H$ acquires its non-zero vev. The vev of $H$ in the true minimum is governed by the fourth term in Eq.~\eqref{eq_W}. In a realistic scenario, we would expect inflation to end by a phase transition, i.e. the $H$ field to be a non-singlet under some symmetry group \footnote{In this case, the terms $\hat{H}^2$ might be replaced by a different combination of fields, i.e.\ by $\hat{H}_1\hat{H}_2$, in order to form a singlet under the symmetry group. The symmetry could, e.g., be a unifying gauge symmetry or a family symmetry.}. A realization within the context of GUTs can be found in Ref.~\cite{Antusch:2010va}, and the idea of relating the ``waterfall'' of hybrid inflation to the breaking of a family symmetry was discussed in Ref.~\cite{Antusch:2008gw}. For simplicity, we keep $\hat{H}$ a gauge singlet here. Finally, the ellipsis represent possible higher dimensional operators. We note that a superpotential of the form given in Eq.~\eqref{eq_W} can be realized in an explicit model with discrete symmetries, as has been shown in~\cite{Antusch:2009ef, Antusch:2010va}. These discrete symmetries distinguish between the gauge singlet fields of Eq.~\eqref{eq_W}.

The parameters which appear in Eq.~\eqref{eq_W}, and which govern our framework, can be understood as follows:
\begin{itemize}
\item The \textbf{phase transition scale} $M$ is the vev of the scalar component of the $\hat{H}$ superfield after the phase transition ending inflation and is the mass scale relevant for inflation.

\item The parameter $\lambda_{11}$ determines the \textbf{seesaw scale} which corresponds to the mass of the lightest right-handed neutrino $m_{N^1} = 2 \frac{\lambda_{11}}{M_P} M^2$ in the true vacuum of the theory.

\item The \textbf{vacuum energy parameter} $\kappa$ fixes the the vacuum energy density \mbox{$V_0 = \kappa^2 M^4$} with regard to the phase transition scale.

\item The \textbf{effective Yukawa coupling} $\tilde{y}_1 \equiv \sqrt{(y_{\nu} y_{\nu}^{\dagger})_{11}}$ is linked to the light neutrino masses.
\end{itemize}

We will be working in a SUGRA framework with SUGRA corrections stabilizing the scalar components of $\hat{S}$, $\hat{L}$ and $\hat{h}$ during inflation and with a symmetry in the K\"ahler potential that guarantees tree-level flat $N^i$ directions. 
This solves the $\eta$-problem of SUGRA inflation. The details of such a SUGRA framework are discussed in section~\ref{sugra}. However, to illustrate the underlying physics more clearly we first focus on a global supersymmetry (SUSY) model and take the features mentioned above for granted.

In order to produce the CP-violation necessary for leptogenesis we work with three (s)neutrino generations. Assuming that the right-handed neutrinos are strongly hierarchical, i.e.\ one of them is significantly lighter than the other two, the scalar components of the latter superfields can be stabilized at their minima before the final 60 e-folds of inflation begin. Thus the time evolution of the lightest sneutrino controls the relevant slow-roll dynamics and it can therefore be identified as the inflaton. On the other hand, the outcome of leptogenesis is governed by the sneutrino with the smallest decay rate. This implies a comparatively small mass and small Yukawa couplings. In the following, we shall concentrate on the case where the lightest sneutrino drives both inflation and leptogenesis. Hence, the three generation model can be simplified to an effective one generation model in the right-handed neutrino sector, with the only remnant of the other two generations being a non-vanishing CP-asymmetry necessary for leptogenesis. We can thus concentrate on $i = 1$ in Eqs.~\eqref{eq_W} and we denote the relevant inflaton direction by $N \equiv N^1$ and the respective coupling constant by $\lambda \equiv \lambda_{11}$.

\section{Inflation \label{inflation}}
Based on the framework described in the previous section, we now have a closer look at the inflationary dynamics in our model. 
Furthermore, we derive restrictions on the model parameters from the requirement of successful inflation and the latest observational data. 
We start with a short introduction to slow-roll inflation and then discuss a realization of the model of 
section~\ref{framework} in a globally supersymmetric context. We then refine this discussion by including SUGRA effects and close the section by listing the inflationary predictions from our model and comparing them to the latest observational data.

\subsection{Short Overview \label{overview}} 
A common way to realize inflation is the so-called slow-roll paradigm, where a classical scalar field with a strong dominance of its potential energy over its kinetic energy $V \gg \mathcal{L}_{\text{kin}}$ drives the accelerated expansion of the universe.
At the same time, the quantum fluctuations of the inflaton field can account for the metric perturbations which give rise to
the small scale CMB anisotropies. Inflation ends when the slow-roll conditions are violated, i.e.~when the slow-roll parameters parameterizing the scalar potential and its derivatives
\begin{equation}
\label{slow_roll_pars}
\epsilon = \frac{M_P^2}{2}\, \left( \frac{V'}{V} \right)^2, \qquad 
 \eta  = M_P^2 \,\frac{V''}{V}  \,, \qquad 
\xi^2 = M_P^4 \,\frac{V' V'''}{V^2} \,,
\end{equation}
become of order one. Here, a prime denotes derivative w.r.t.~the inflaton field $N$.

In the slow-roll approximation, when $\epsilon \ll 1$, $\left| \eta \right| \ll 1$ and $\xi^2\ll1$, the equation of motion of a homogeneous (classical) scalar field
\begin{equation}\label{eom_scalar}
\ddot{N} + 3\,\mathcal{H}\, \dot{N} + V' = 0\,,
\end{equation}
simplifies to
\begin{equation}\label{eomI}
3 \,\mathcal{H} \,\dot{N} = -V'\,.
\end{equation}
Here $\cal H$ denotes the Hubble expansion parameter.

Models of inflation typically predict the power spectra of the gauge invariant scalar and tensor perturbations at the time when the relevant fluctuations exited the horizon, roughly $\mathcal{N}_{\text{e}}\simeq 50 - 70$ e-folds before the end of inflation. The amplitude of the scalar perturbations $\Delta_s^2$, the scalar spectral index $n_s$, the running of the scalar spectral index $\alpha_s$, the tensor-to-scalar ratio $r$ and the tensor spectral index $n_t$ can be estimated in terms of the potential and the slow-roll parameters~\cite{Baumann:2009ds} as
\begin{equation}
\label{eq_infl_pred}
\begin{split}
\Delta_s^2 &\simeq \frac{1}{M_P^6}\,\frac{1}{12 \pi^2} \,\frac{V^3}{(V')^2}\,,\\
n_s &\simeq 1 - 6\, \epsilon + 2 \,\eta\,, \\ 
\alpha_s &\simeq 16 \,\epsilon\, \eta - 24 \,\epsilon^2 - 2\, \xi^2\,,\\
r &\simeq 16 \,\epsilon \,,\\
n_t &\simeq - 2\, \epsilon\,,
\end{split}
\end{equation}
where these expressions have to be evaluated at the field value $N = N(\mathcal{N}_{\text{e}})$ when the relevant scales leave the horizon. This value can be computed from Eq.~\eqref{eomI}.

In order to test our model against observations, we compare the predictions from Eqs.~\eqref{eq_infl_pred} to the experimental data obtained from the 7 year WMAP survey combined with measurements of the baryon acoustic oscillations (BAO)~\cite{Percival} and measurements of the present value of the Hubble parameter $H_0$~\cite{Riess} using the six parameter $\Lambda$CDM fit~\cite{wmap}, which are given by
\begin{equation}
\label{infl_obs}
\begin{split}
&0.951 < n_s < 0.975 \qquad \text{(68 $\%$ CL)}\,,\\
&0.939 < n_s < 0.987 \qquad \text{(95 $\%$ CL)} \,,\\
&\Delta_s^2 = (2.441^{+0.088}_{-0.092}) \cdot 10^{-9}\,.
\end{split}
\end{equation}

\subsection{Realization in Global Supersymmetry \label{susy}}

In the model described in section~\ref{framework}, we assumed a symmetry in the K\"ahler potential guaranteeing a flat $N$ direction (imaginary direction of the scalar component of $\hat{N}^1$) at tree-level (see also section~\ref{sugra}). Thus loop corrections must be taken into account and these can indeed generate a small slope as required for slow-roll inflation. According to~\cite{coleman_weinberg_potential}, the Coleman-Weinberg one-loop effective potential is given by
\begin{equation}
\label{vloop}
V_{\text{loop}} = \frac{1}{64 \pi^2} \, \text{STr} \left[ {\cal M}^4 \left( \ln \frac{{\cal M}^2}{Q^2} - \frac{3}{2}\right) \right] \,,
\end{equation}
with ${\cal M}$ denoting the mass matrix of the theory and $Q$ a renormalization scale. The $N$-dependent bosonic and fermionic mass terms generating a slope for the inflaton via the loop potential can be calculated from the scalar F-term potential
\begin{equation}
\label{global_scalar_pot}
V_F = \sum_i \left| \; \frac{\delta W(\hat{\Phi})}{\delta \hat{\Phi}^i} \, \bigg|_{\hat{\Phi} \rightarrow \Phi} \; \right|^2 \,,
\end{equation}
and the fermionic mass matrix
\begin{equation}
({\cal M}_F)_{ij} = \frac{\delta^2 W(\hat{\Phi})}{\delta \hat{\Phi}^i \delta \hat{\Phi}^j} \, \bigg|_{\hat{\Phi} \rightarrow \Phi}\,.
\end{equation}
$\hat{\Phi}^i$ denote the superfields of the theory, $\Phi^i$ the respective scalar components. The relevant (i.e. $N$-dependent) contributions to the loop potential are give by the $\hat{H}, \, \hat{L}^j$ and $\hat{h}$ mass terms \footnote{Here the index (S) (for scalar) denotes mass terms of the real parts of the complex spin-0 components of the superfields whereas the index (P) (for pseudoscalar) marks the mass terms corresponding to the purely imaginary parts. The index (F) marks the mass terms of the fermionic components of the superfields.}
\begin{equation}
\label{eq_global_mass}
\begin{split}
(m_{h_{ a}}^{(S)})^2 &= (m_{h_{ a}}^{(P)})^2 = (m_{h_{ a}}^{(F)})^2 = \frac{1}{2} N^2 \sum_j |(y_{\nu})_{1j}|^2\,, \\
(m_{L^j_a}^{(S)})^2 &= (m_{L^j_a}^{(P)})^2 = (m_{L^j_a}^{(F)})^2 = \frac{1}{2} N^2 |(y_{\nu})_{1j}|^2\,, \\
(m_H^{(S)})^2 &= 2 \,\kappa^2 M^2\left(x - 1\right) \,, \\
(m_H^{(P)})^2 &= 2 \,\kappa^2 M^2\left(x + 1\right) \,, \\
(m_H^{(F)})^2 &= 2 \,\kappa^2 M^2 x \,,
\end{split}
\end{equation}
with
\begin{eqnarray}
x \equiv \frac{N^4 \lambda^2}{2 \,\kappa^2 M^2 M_P^2 }\,.
\end{eqnarray}
Note that the $\hat{L}^j$ and $\hat{h}$ terms in the supertrace vanish since the degeneracy in the respective fermionic and bosonic masses leads to a cancellation of these contributions. 
Embedding this model in SUGRA provides the necessary stabilization of the scalar components of the $\hat{L}^j$ and $\hat{h}$ superfields during inflation and removes this degeneracy. However in the parameter range of interest, the contribution of the $\hat{L}^j$ and $\hat{h}$ terms to the loop potential turn out to be negligible (see also section \ref{sugra}).
In the following, we fix the renormalization scale to $Q = \sqrt{2} \,\kappa \,M$, which is the order of magnitude of the SUSY breaking scale. In our model, inflation ends when the $H$-field destabilizes at the critical value $N^c$ characterized by $m_H^{(S)} = 0$ and thus $x = 1$,
\begin{equation}
(N^c)^2 = \sqrt{2} \,\frac{\kappa }{\lambda}\, M \, M_P \, .
\end{equation}

We can now determine the observables describing the CMB fluctuations given by Eqs.~\eqref{eq_infl_pred}, thus obtaining expressions depending on the phase transition scale $M$, the seesaw scale $m_N \sim \lambda$ and the vacuum energy parameter $\kappa$. With $x$ as defined above, a Taylor expansion in $1/x$ (with $1/x < 1$ because $N > N^c$ during inflation) yields
\begin{equation}
V_{\text{loop}} \simeq \frac{\kappa^4 M^4}{8\, \pi^2} \, \ln x \, .
\end{equation}
Inserting this into the equation of motion Eq.~\eqref{eomI}, with ${\cal H}$ approximately constant, gives the value for $N$ at ${\cal N}$ e-folds before the end of inflation:
\begin{equation}
\label{eq_nc_global}
N^2({\cal N}) = (N^c)^2 + \frac{{\cal N} \kappa^2}{\pi^2}\, M_P^2\, .
\end{equation}
With this, the inflationary predictions of Eqs.~\eqref{eq_infl_pred} are given by
\begin{equation}
\label{pred_global}
\begin{split}
\Delta_s^2 &\simeq \frac{\pi^2 M^4 N^2}{3\, \kappa^2 M_P^6} \,, \\
n_s &\simeq 1 - \left(1 + \frac{3\kappa^2 }{4 \,\pi^2} \right) \frac{\kappa^2 M_P^2}{\pi^2 N^2}\,, \\
\alpha_s &\simeq - \left(1 + \frac{\kappa^2}{\pi^2} + \frac{ 2 \,\kappa^4}{8\, \pi^4} \right) \frac{\kappa^4 M_P^4}{\pi^4 N^4}\,,  \\
r &\simeq 2\, \frac{\kappa^4 M_P^2}{\pi^4 N^2}\,,  \\
n_t &\simeq - \frac{\kappa^4 M_P^2}{4\, \pi^4 N^2} \,. \\
\end{split}
\end{equation}

\subsection{Embedding in Supergravity}\label{sugra}
We next consider a possible embedding of our model in SUGRA. We focus on a K\"ahler potential with the $\eta$-problem~\cite{Copeland:1994vg, Dine:1995uk} resolved by a shift symmetry~\cite{Antusch:2009ef}
in the inflaton direction
\begin{equation}
\label{eq_K}
K = |\hat{S}|^2 + |\hat{H}|^2 + |\hat{h}|^2 + \sum_i \frac{1}{2}\left(\hat{N}^i+(\hat{N}^i)^{\dagger}\right)^2 + \sum_j |\hat{L}^j|^2 + \frac{\kappa_{SH}}{M_P^2}|\hat{S}|^2|\hat{H}|^2 + \ldots\,.
\end{equation}
The K\"ahler potential can be seen as a general expansion in the superfields of the theory with the additional feature of a shift symmetry which guarantees tree-level flat directions for the imaginary parts of the scalar components of $\hat{N}^i$ and thus possible inflaton directions. The $N$-dependent mass terms generating a slope for the inflaton via the loop potential can be calculated from the scalar F-term potential and the fermionic mass matrix as before by
\begin{equation}
\label{sugrapotential}
\begin{split}
V_F &= e^K \left[ K^{i\overline{j}}D_i W D_{\overline{j}} W^* - 3|W|^2 \right] \; \Big|_{\hat{\Phi} \rightarrow \Phi} \,, \\
({\cal M}_F)_{ij} &= e^{K/2}(W_{ij} + K_{ij}W + K_iW_j + K_j W_i + K_i K_j W - K^{k\overline{l}}K_{ij\overline{l}} {\cal D}_k W) \; \Big|_{\hat{\Phi} \rightarrow \Phi} \,.
\end{split}
\end{equation}
The relevant contributions to the loop potential are given by the $\hat{H}, \, \hat{L}^j$ and $\hat{h}$ mass terms which now obtain SUGRA corrections:
\begin{equation}
\begin{split}
(m_H^{(S)})^2 &= 2\, \kappa^2 M^2\left[ x - 1 +\left(\frac{M}{M_P}\right)^2 \left(1 - \kappa_{SH}\right)/2\right] \,,\\
(m_H^{(P)})^2 &= 2 \,\kappa^2 M^2\left[ x + 1 + \left(\frac{M}{M_P}\right)^2 \left(1 - \kappa_{SH}\right)/2\right] \,,\\
(m_H^{(F)})^2 &= 2 \,\kappa^2 M^2 x \,,\\
(m_{h_{ a}}^{(S)})^2 &= (m_{h_{ a}}^{(P)})^2 =  \kappa^2\frac{M^4}{M_P^2} + \frac{1}{2}\, N^2 \sum_j |(y_{\nu})_{1j}|^2\,,\\
(m_{h_{ a}}^{(F)})^2 &= \frac{1}{2}\, N^2 \sum_j |(y_{\nu})_{1j}|^2\,, \\
(m_{L^j_a}^{(S)})^2 &= (m_{L^j_a}^{(P)})^2 = \kappa^2\frac{ M^4}{M_P^2} + \frac{1}{2}\, N^2 |(y_{\nu})_{1j}|^2\,, \\
(m_{L^j_a}^{(F)})^2 &= \frac{1}{2}\, N^2 |(y_{\nu})_{1j}|^2 \,.
\end{split}
\end{equation}
Comparing these expressions to the mass terms calculated in section~\ref{susy} we note some important points. A second mass scale, the scale of the SUGRA mass splitting $\kappa M^2 / M_P$, has appeared. However this scale is much smaller than the SUSY mass splitting scale $ \sqrt{2}\, \kappa \,M$ and thus we shall keep the latter scale as the renormalization scale. The additional mass splitting implies that the $\hat{h}$ and $\hat{L}^j$ contributions no longer cancel. However since the mass splitting is small compared to the SUSY mass splitting appearing in the $\hat{H}$ mass terms and since the remaining parts of the $\hat{h}$ and $\hat{L}^j$ mass terms are proportional to $|(y_{\nu})_{1j}|$ these additional contributions to the loop potential are negligible for $\tilde{y}_1 < 10^{-2}$. We will see later that this easily holds in our model. Furthermore, a new parameter has appeared in the loop potential:
\begin{itemize}
\item The \textbf{SUGRA correction parameter} $\kappa_{SH}$ controls the SUGRA corrections to the loop potential. $\kappa_{SH}= 1$ recovers the phenomenology of global SUSY.
\end{itemize}

\subsection{Predictions \label{predictions}}

\begin{figure}
\subfigure[]{\includegraphics[width=0.485\textwidth]{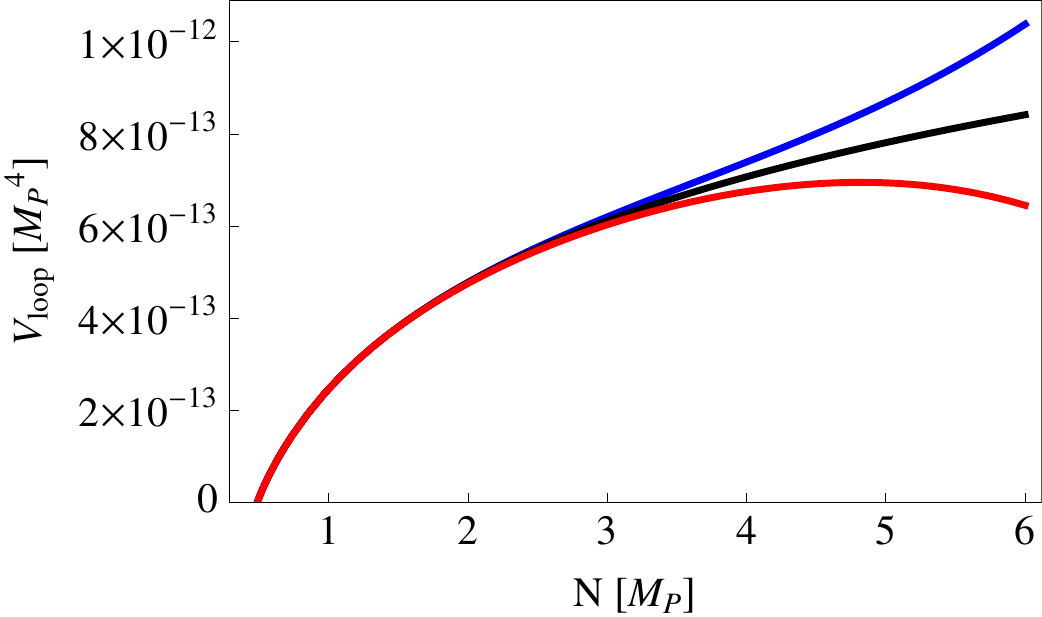}} \hfill \hspace{0.3cm}
\subfigure[]{\includegraphics[width=0.49\textwidth]{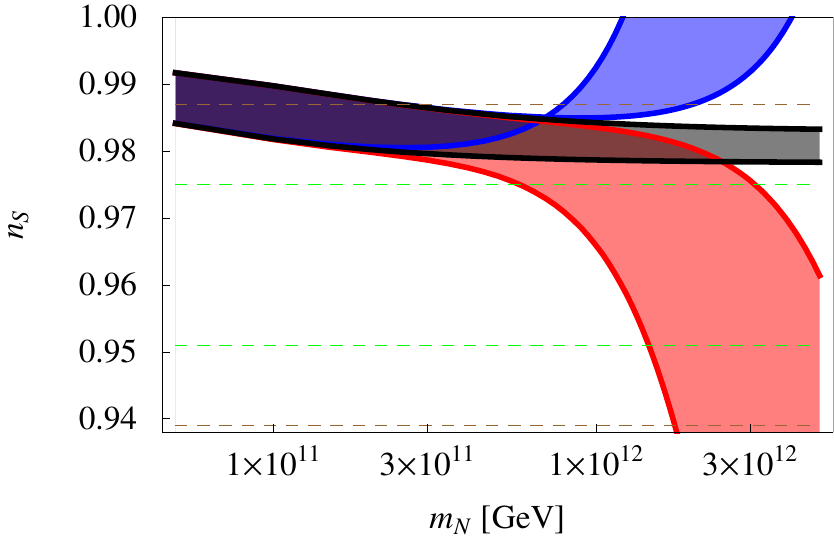}}
\subfigure[]{\includegraphics[width=0.49\textwidth]{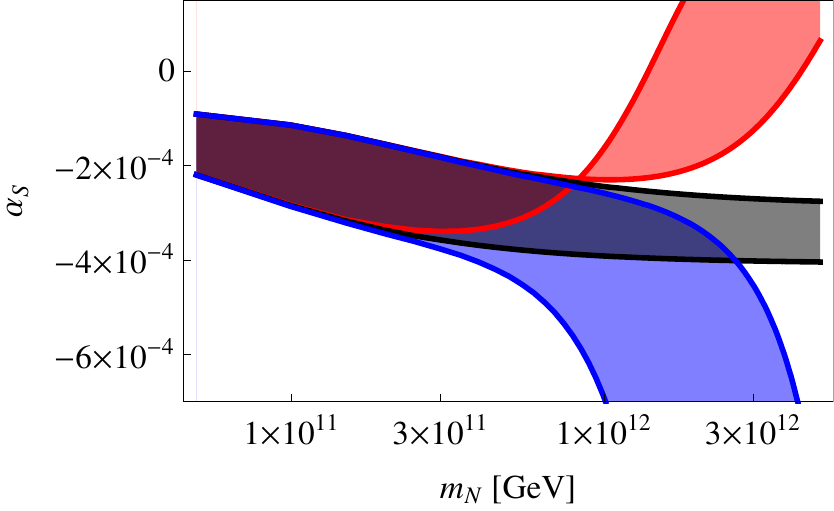}} \hfill \hspace{0.3cm}
\subfigure[]{\includegraphics[width=0.49\textwidth]{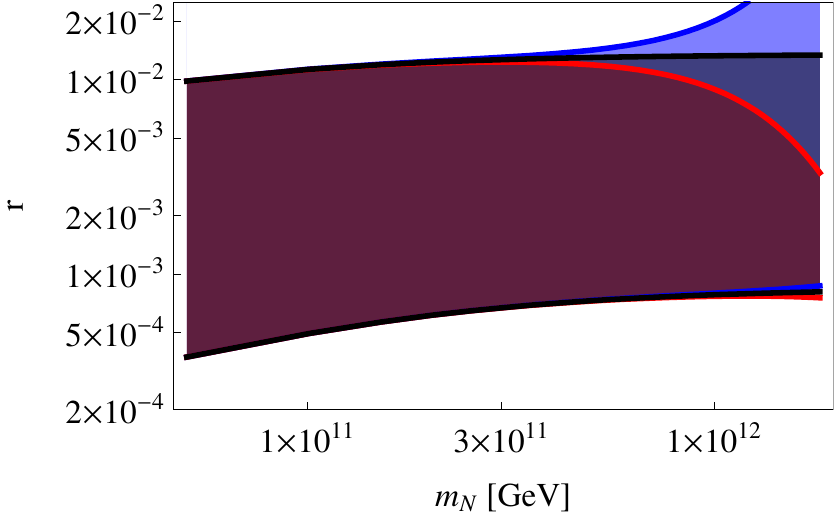}}
\caption{Loop potential and predictions for CMB observables for $\kappa_{SH} - 1= -1,\,0 \, \text{and} \,1$ (blue, black and red). (a) Loop potential for $\kappa = 0.5, \, M = 0.0032 \, M_P, \, m_N = 2.5 \cdot 10^{11}$ GeV. (b) - (d): Spectral index, running of spectral index and tensor-to-scalar ratio. The width of the bands is given by the variation of the vacuum energy parameter $\kappa = 0.5 - 2$. On the left border of plots (b) and (c), $\kappa = 0.5$ corresponds to the upper set of lines, and in plot (d) to the lower set of lines. For labeling the x-axis, the phase transition scale was set to $M = 0.0032 \, M_P$. The 95$\%$ and 68$\%$ CL  experimental bounds from Eqs.~\eqref{infl_obs} are marked by dashed lines in (b).}
\label{fig_pred}
\end{figure}

As in the globally supersymmetric case, predictions for observables describing the CMB spectrum can now be obtained by solving Eq.~\eqref{eomI} and evaluating Eqs.~\eqref{eq_infl_pred} at the time when the CMB fluctuations exited the horizon. In the SUGRA scenario, this was done numerically for ${\cal N}_e = 60$. Fixing the phase transition scale $M$ by the experimental value for the amplitude of the CMB fluctuations $\Delta_s^2$, the behavior of the spectral index $n_s$, its running $\alpha_s$ and the tensor-to-scalar ratio $r$ is shown in Fig.~\ref{fig_pred}.
Interpreting the results visualized in Fig.~\ref{fig_pred} and enforcing the experimental bounds of Eqs.~\eqref{infl_obs} implies restrictions on the model parameters.
\begin{itemize}

\item The \textbf{phase transition scale $M$} is fixed to $M \simeq 0.0032 \, M_P \simeq 8 \cdot 10^{15}$ GeV with a slight deviation in the region of large SUGRA corrections. This is consistent with the global SUSY calculation (from Eqs.~\eqref{eq_nc_global} and~\eqref{pred_global}) which gives $M^4 \simeq 3 \Delta_s^2/{\mathcal{N}_{\text{e}}}$ for $N(\mathcal{N}_{\text{e}}) \gg N^c$.

\item The width of the band in Fig.~\ref{fig_pred} is given by the variation of the \textbf{vacuum energy parameter $\kappa$}. A priori we would expect $\kappa$ to be an ${\cal O}(1)$ parameter, thus we shall assume $0.5 < \kappa < 2$. In Fig.~\ref{fig_pred}, larger values of $\kappa$ are associated with SUGRA corrections becoming relevant at smaller values of $m_N$. In particular the tensor-to-scalar ratio $r$ is quite sensitive to $\kappa$ with $\kappa = 2$ leading to comparatively larger $r \sim {\cal O}(10^{-2})$.

\item Fig.~\ref{fig_pred} also demonstrates the effect of the \textbf{SUGRA correction parameter} $\kappa_{SH}$. The respective quantities are marked in black for $\kappa_{SH} -1 = 0$ which corresponds to the globally supersymmetric limit and in blue (red) for $\kappa_{SH} -1 = -1 \, (+1)$ which corresponds to turning on the SUGRA corrections in the $\hat{H}$ mass terms with positive (negative) sign. In the considered SUGRA context the value of $\kappa_{SH}$ is a priori undetermined. Thus we would in general not expect to find global SUSY restored, which would correspond to 
$\kappa_{SH}$ exactly equal to one.

\item The second parameter controlling the effect of the SUGRA corrections is the \textbf{seesaw scale} $m_N = 2 \,\lambda\, M^2/M_P$. Fig.~\ref{fig_pred} shows that these corrections are suppressed for small $m_N$, i.e.\ the observables are independent of $\kappa_{SH} $ for small values of the seesaw scale and the model predicts (for $M$ fixed by the experimental value of~$\Delta_s^2$)
\begin{equation}
0.98 < n_s < 1 \,,\qquad
3 \cdot 10^{-4} < \alpha_s < 0 \,, \qquad
r < 0.013\,.
\end{equation}
Note that in this case Eqs.~\eqref{pred_global} hold. For very small values of the seesaw scale $m_N$, the spectral index $n_s$ approaches 1, which is not preferred by the latest WMAP data. On the other hand, all solutions with $\kappa_{SH} \neq 1$ leave the experimentally preferred region for the spectral index at large values of $m_N$. 
In combination, we find the preferred regions $2 \cdot 10^{10}\, \text{GeV} \lesssim m_N \lesssim 7 \cdot 10^{12}\, \text{GeV}$ for $\kappa_{SH}-1= + 1$ and $2 \cdot 10^{10}\, \text{GeV} \lesssim m_N \lesssim 2 \cdot 10^{12}\, \text{GeV}$ for $\kappa_{SH} -1= -1$, respectively. \footnote{Equivalently, we obtain $5 \cdot 10^{-4} < \lambda < 0.13 (0.043)$ for $\kappa_{SH}-1 = + 1$ and $\kappa_{SH} -1 = -1$, respectively.}
\end{itemize}
Motivated by the above results, we take the phase transition scale to be fixed at $M \simeq 8 \cdot 10^{15} \, \text{GeV}$ and concentrate on the parameter ranges
\begin{equation}
\label{eq_infl_preferred}
2 \cdot 10^{10}\, \text{GeV} < m_N < 7 \cdot 10^{12} \, \text{GeV} \,, \qquad
0.5 < \kappa < 2 \,, \qquad
|\kappa_{SH} - 1| > 0.1  \,,
\end{equation}
in the further discussion. This yields
\begin{equation}
-0.0004 \lesssim \alpha_s \lesssim 0.0002\, , \qquad r \lesssim 0.015 \,,
\end{equation}
for the running of the spectral index and the tensor-to-scalar ratio.

\section{Reheating and Leptogenesis}\label{leptogenesis}

After the end of the inflationary epoch the homogeneous classical fields and their quantum fluctuations evolve according to their respective equations of motion. The universe enters a matter dominated regime until the decay of heavy particles and the thermalization of the light particles result in the total energy density being dominated by radiation 
(Fig.~\ref{fig_numerics}). This out-of-equilibrium decay of heavy particles can furthermore produce the necessary lepton asymmetry. In the following we will study these processes for the classical fields (reheating) and briefly comment on possible effects originating from their fluctuations (preheating). To this end we will start with the equations of motion for classical scalar fields, justify a simplification to Boltzmann equations and finally derive analytical expressions for the generated baryon asymmetry and the reheat temperature. We finish by commenting on preheating via parametric resonance and tachyonic preheating in this context.

\subsection{Classical Field Dynamics after Inflation}\label{Classical Field Dynamics after Inflation}
The equations of motion for the scalar fields can be obtained by adding a phenomenological decay term~\cite{Kofman:1997yn} to Eq.~\eqref{eom_scalar} thus giving
\begin{equation}
 \ddot{\phi} + 3\, {\cal H} \,\dot{\phi} + V'(\phi) + \Gamma \dot{\phi} = 0 \quad \text{with} \quad \phi =\{N,\, H\}\, .
\end{equation}
Adding a Boltzmann equation for the quickly thermalizing\footnote{In principle, thermalization in the MSSM could be delayed if MSSM flat directions obtain large vevs (see e.g.~\cite{Allahverdi:2005mz, Allahverdi:2007zz}). However in our scenario, this is not the case since only the $N$ direction is protected against large SUGRA corrections.} ultra-relativistic particles and the Friedmann equation, we arrive at a closed set of differential equations:
\begin{align}
\label{n}
\ddot{N} + 3\, {\cal H} \,\dot{N} + \frac{\partial V}{\partial N} + \Gamma_N \dot{N}  &= 0\,, \\
\label{eq h}
\ddot{H} + 3\, {\cal H}\, \dot{H} + \frac{\partial V}{\partial H} + \Gamma_H \dot{H} &= 0\, ,\\
\label{rrad}
\dot{\rho_{R}} + 4\, {\cal H}\, \rho_{R} - \Gamma_N \rho_N - \Gamma_H \rho_H  &= 0\,,\\
\label{friedmann}
\frac{1}{3\,M_P^2}\left(\rho_N + \rho_H + \rho_R\right) &= {\cal H}^2  \,,
\end{align}
with $\rho_R$, $\rho_N$ and $\rho_H$  denoting the energy densities of the ultra-relativistic particles, the $N$-field and the $H$-field respectively with $\rho_{N} + \rho_H = \dot{N}^2/2 + \dot{H}^2/2 + V(N,H)$. Having solved Eq.~\eqref{n}- \eqref{friedmann}, the lepton number density $n_L$ can be calculated from the Boltzmann equation
\begin{equation}
\label{n asy}
 \dot{n_L} + 3\, {\cal H}\, n_L = \epsilon_{1}  \Gamma_N \frac{\rho_N}{m_{N_1}} + \epsilon_{3} \Gamma_H \frac{\rho_H}{m_{N_3}} \,, 
\end{equation}
with the CP-violation per (s)neutrino decay $\epsilon$ for a hierarchical spectrum of right-handed neutrinos bounded by~\cite{Covi:1996wh, Davidson:2002qv, Hamaguchi:2001gw}
\begin{equation}
\label{epsilon}
 \epsilon_{i}  < \frac{3}{8 \pi} \frac{\sqrt{\Delta m^2_{\text{atm}}} m_{N_i}}{ \langle v \rangle ^2} \,.
\end{equation}
Here $\langle v \rangle$ denotes the vacuum expectation value of the up-type Higgs. The lepton asymmetry is typically normalized to the entropy density $s = 2 \pi^2 g_* T^3 / 45$~\cite{Bailin:2004zd} with the effective number of degrees of freedom $g_* = 915/4$ for the MSSM particles. The asymmetry $n_L / s$  is transferred to the baryon sector via sphaleron processes $n_B~=~\frac{C}{C - 1} n_L$ where $C$ is a number ${\cal O}(1)$ depending on the field content of the model and the temperature $T_{sph}$ when the sphalerons leave equilibrium. In the MSSM $C \sim 1/3$~\cite{Davidson:2008bu}. The quantity measured today is $\eta \equiv \frac{n_B}{n_{\gamma}}$ which can be calculated from the results above using the current conversion factor 
$s = 7.04 \, n_{\gamma}$~\cite{Bailin:2004zd}. The second important physical quantity in the theory of reheating is the temperature of the universe when the universe becomes radiation dominated ($\Gamma_N \approx {\cal H}$), the so-called reheat temperature. It can be calculated from the results above using $T^4 = \rho_R \cdot 30 / (g_* \pi^2)$.

Eqs.~\eqref{n} - \eqref{n asy} assume that both the $N$-particles and the $H$-particles decay into ultra-relativistic particles with the respective decay rates $\Gamma_N$ and $\Gamma_H$. We will now describe a possibility how to evaluate these quantities in our framework. At the beginning of the reheating phase ${\cal H} \gg \Gamma_N, \Gamma_H$ holds which implies that the decaying particles are damped predominantly by Hubble expansion, not by decays, and the produced ultra-relativistic particles are strongly diluted. The decays become significant for $t \sim {\cal H}^{-1} \sim \text{min} \{ \Gamma_N^{-1}, \Gamma_H^{-1} \}$. At this stage it is safe to assume $N \ll 1$. In this limit the respective decay rates derived from Eq.~\eqref{eq_W} are
\begin{align}
 \Gamma_N &\simeq \lim_{N \rightarrow 0} \frac{(y_{\nu} y_{\nu}^{\dagger})_{11}}{4 \pi} m_N = \frac{(y_{\nu} y_{\nu}^{\dagger})_{11}}{2 \pi} \frac{\lambda}{M_P} M^2 \,, \\
\Gamma_H &\simeq \lim_{N \rightarrow 0} \min\{ \Gamma_{H \rightarrow N_3 N_3}, \Gamma_{N_3 \rightarrow hL} \} = \frac{2}{\pi} \frac{\lambda_{33}}{M_P} M^2 \min \{ (y_{\nu} y_{\nu}^{\dagger})_{33}, \frac{\lambda_{33} \kappa M}{16 M_P} \} \,, 
\end{align}
with the sneutrino $N$ decaying directly into lepton and Higgsino or slepton and Higgs and the $H$ particles decaying predominantly into the heaviest fermionic neutrino (assuming this is not strongly suppressed by kinematics)
\footnote{Note that $\lim_{N \rightarrow 0}  \Gamma_{H \rightarrow N_i N_i} \simeq \frac{\lambda_{ii}^2 \kappa M^3}{8 \pi M_P^2} (1 - 2\frac{\lambda_{ii}^2 M^2}{\kappa^2 M_P^2}) (1 - 4 \frac{\lambda_{ii}^2 M^2}{\kappa^2 M_P^2})^{1/4}$. The expressions in brackets emphasize that the decay is kinematically possible if $m_H^{(S)} > 2 m_{N_i}^{(F)}$. However in the parameter range of interest, $\lambda_{ii}/M_P \ll \kappa/M$ holds, as can be seen from Eqs.~\eqref{eq_infl_preferred}. Thus we have $\Gamma_{H \rightarrow N_i N_i} \simeq \frac{\lambda_{ii}^2 \kappa M^3}{8 \pi M_P^2}$.}, which then in turn decays into lepton and Higgs or slepton and Higgsino.

Note that the Boltzmann equations~\eqref{rrad} and~\eqref{n asy} imply a splitting of the total matter energy density $\rho_M$ into $\rho_N$ and $\rho_H$, which is not straightforward if the respective degrees of freedom are highly coupled. However, since $\Gamma_N \ll \Gamma_H$ in our setting \footnote{The assumption $\Gamma_{N_1} < \Gamma_{N_3}$ (see section~\ref{framework}) implies $\lambda_{33} (y_{\nu} y_{\nu}^{\dagger})_{33} > \lambda_{11} (y_{\nu} y_{\nu}^{\dagger})_{11}$. The assumption of hierarchical heavy neutrinos implies $\lambda_{33} \gg \lambda_{11}$.} in the preferred region of parameter space (see section~\ref{infl_lepto}), any radiation energy density produced by $H$-decays will be strongly diluted during the following matter dominated phase governed by oscillations of the sneutrino. With $\rho_M \approx \rho_N$ shortly after the end of inflation due to the strong damping of the $H$-field (see below) we can thus substitute~\eqref{rrad} and~\eqref{n asy} by
\begin{align}
 \label{rrad2}
&\dot{\rho_{R}} + 4 \,{\cal H}\, \rho_{R} - \Gamma_N \rho_M \simeq 0 \,, \\
\label{n asy 2}
 &\dot{n_L} + 3\, {\cal H}\, n_L \simeq \epsilon_{1}  \Gamma_N \frac{\rho_M}{m_{N_1}} \,, 
\end{align}
without introducing a significant error for the finally produced radiation density. The effect of this approximation on the Hubble expansion rate is negligible since the Friedmann equation is predominantly governed by $\rho_M$ for $t < \Gamma_N^{-1}$.

\begin{figure}
\subfigure[]{\includegraphics[width=0.49\textwidth]{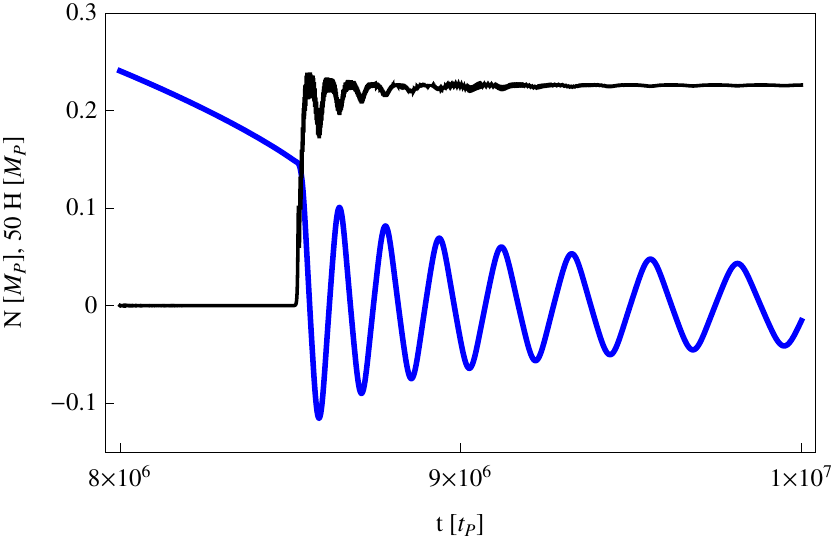}}\hfill
\subfigure[]{\includegraphics[width=0.49\textwidth]{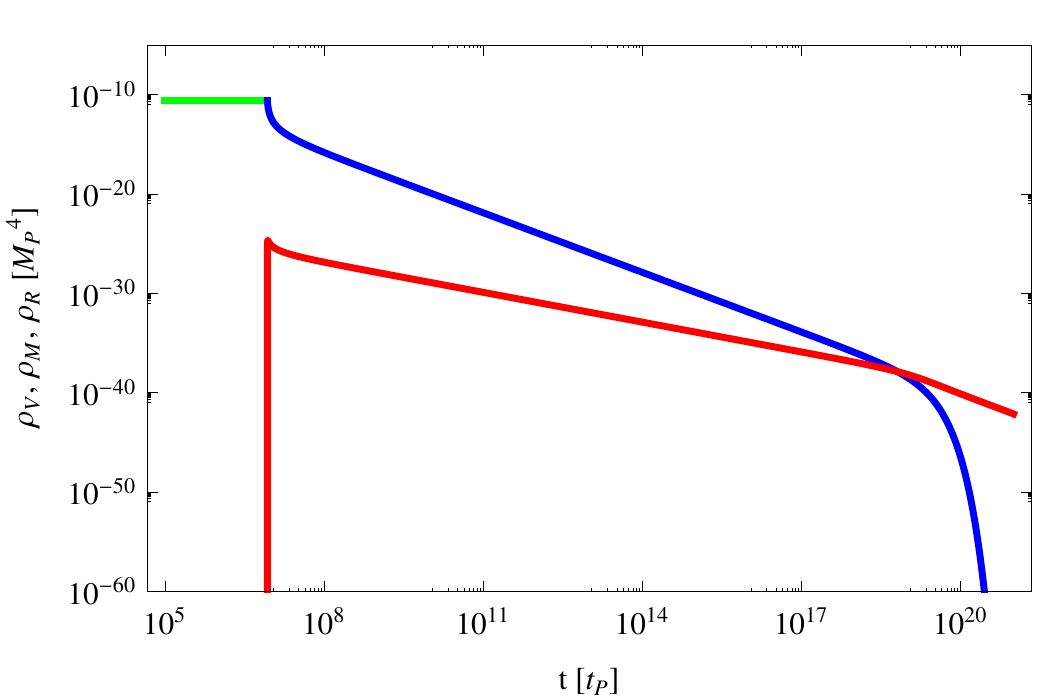}}
\caption{(a) Field dynamics of the sneutrino field (blue) and the waterfall field (black). (b) Evolution of the vacuum (green), matter (blue) and radiation (red) energy densities. The parameters chosen for these plots are $\kappa = 0.5$, $M = 0.0032 \ M_P$, $m_N = 4.9 \cdot 10^{12}$ GeV, $\tilde{y}_1 = 10^{-6}$, $\kappa_{SH} = 0.5$. The timescale is given in units of the Planck time, $t_P = \hbar/M_P \simeq 2.7 \cdot 10^{-43} s$.}
\label{fig_numerics}
\end{figure}

Solving Eqs.~\eqref{n},~\eqref{eq h},~\eqref{friedmann},~\eqref{rrad2} and~\eqref{n asy 2} numerically, we obtain the time evolution of scalar fields, the energy densities, the scale factor and the lepton asymmetry, respectively. The former two are displayed in Fig.~\ref{fig_numerics}. The regime of reheating is characterized by oscillating scalar fields and can be divided into distinct phases: After the end of inflation both $N$ and $H$  fall to their true minimum and begin to oscillate. After only a few oscillations the classical field $H$ settles at its minimum and the dynamics of the system is governed by the oscillation of the $N$ field. The further evolution of the $N$ oscillations is governed by Hubble damping. As long as ${\cal H} \gg \Gamma_N$ the universe is governed by (damped) oscillating scalar fields which can be interpreted as (decaying) heavy particles. This implies a matter dominated universe out of thermal equilibrium. Ultra-relativistic particles are produced through the decays of the heavy particles, however they are diluted by the expansion of the universe. As soon as ${\cal H} \approx \Gamma_N$ the radiation energy density becomes dominant and the light particles begin to thermalize. This marks the end of reheating and determines the reheat temperature and the asymmetry $n_L/s$.

\subsection{Simplified Treatment with Boltzmann Equations}\label{Simplified Treatment with Boltzmann Equations}
Since the set of equations~\eqref{n},~\eqref{eq h},~\eqref{friedmann} and~\eqref{rrad2} is quite involved, a common attempt in the literature (e.g.~\cite{Kolb:1990vq}) is to simplify these equations by time-averaging the equations of motion of the scalar fields. The result is a set of Boltzmann equations for the matter energy density $\rho_M = \rho_N + \rho_H$ and the radiation energy density $\rho_R$ completed by the Friedmann equation
\begin{align}
\label{rhon}
\dot{\rho_{M}} + 3\, {\cal H}\, \rho_{M} &= -\Gamma_N \rho_{M}, \\
\dot{\rho_R} + 4\, {\cal H}\, \rho_R &= \Gamma_N \rho_{M},  \\
\label{H}
\frac{1}{3\, M^2_P}\left(\rho_M + \rho_R\right) &=  {\cal H}^2 .
\end{align}
The lepton asymmetry is determined by Eq.~\eqref{n asy 2}. The big advantage is that these equations have approximate analytical solutions. However their derivation (see e.g.~\cite{Kolb:1990vq}) implies an important assumption concerning the scalar potential $V(N,H)$. In order to rewrite the time-averaged kinetic energy density in terms of the total energy density by exploiting the Virial theorem we must assume that we can write the scalar potential as $V(N,H) = V_N(N) + V_H(H)$ with $V_N \sim N^r$ and $V_H \sim H^r$. Eqs.~\eqref{rhon} - \eqref{H} are obtained with $r = 2$. Numerical simulations of the full system~\eqref{n},~\eqref{eq h},~\eqref{friedmann},~\eqref{rrad2} and~\eqref{n asy 2} show that this assumption is not justified in the early oscillation phase in the model described by Eq.~\eqref{eq_W} since the large oscillations of the $N$-field result in a highly coupled system with higher orders terms in the scalar potential playing a non-negligible role. However they do also show that for $t \approx \Gamma^{-1} \approx {\cal H}^{-1}$ the simpler system of differential equations~\eqref{rhon} - \eqref{H} does give a good approximation. This is the point of time relevant for the predictions of the reheating phase.

Having seen that the results of the numerical solutions to the full field equations for $t \approx \Gamma^{-1}$ can be approximated reasonably well by the simpler set of Boltzmann differential equations, we can now find approximate analytical solutions to the latter and use these expressions to find estimates for the reheat temperature and the produced baryon asymmetry
\begin{align}
\label{Trh}
 T_{RH} &\approx \left(\frac{9}{4 \pi^{4} g_*}\right)^{1/4} \sqrt{(y_{\nu} y_{\nu}^{\dagger})_{11} m_N M_P}\,, \\
\label{nB}
 \frac{n_B}{n_{\gamma}} (t_0)  &\approx 3.45 \,\frac{C}{C-1}  \,  g_*^{-1/4}  \epsilon\, \sqrt{\frac{(y_{\nu} y_{\nu}^{\dagger})_{11}}{m_N/M_P}}\,.
\end{align}
Combining~\eqref{Trh} and~\eqref{nB} reproduces the familiar relation $n_B/s \sim \epsilon\ T_{RH}/ m_N$ (see e.g.~\cite{Bailin:2004zd, Kolb:1990vq}). These results must be compared with existing bounds on the reheating process. The WMAP 7 year data combined with measurements of the baryon acoustic oscillations and todays Hubble parameter imply $\frac{n_B}{n_{\gamma}} = (6.19 \pm 0.15)  \cdot 10^{-10}$ \cite{Komatsu:2010fb}, thus yielding $T_{RH} > 1.4 \cdot 10^6$ GeV.

Additionally, the reheat temperature is bounded from above by the so-called gravitino problem~\cite{Khlopov:1984pf, Ellis:1984eq, Ellis:1984er, Moroi:1993mb, Kawasaki:2004yh}. A high reheat temperature would result in an overproduction of gravitinos. If these are stable, then the fact that their energy density can not be larger than the present total energy density of the universe leads to a bound on the reheat temperature in terms of the gravitino mass $m_{3/2}$. On the other hand, if gravitinos are not stable, they can either decay before or during and after the Big Bang Nucleosynthesis (BBN). In the former case (i.e.~heavy gravitinos), with R-parity conserved the gravitinos will decay into the lightest supersymmetric particle (LSP) and their production is thus constrained by the dark matter abundance. This yields a fairly model independent bound of $T_{RH} < 2 \cdot 10^{10}$ GeV for an LSP mass of about 100 GeV to 150 GeV. In the latter case (i.e.~light gravitinos), the decay of the gravitinos would alter the outcome of BBN and create a conflict between BBN predictions and observations. This yields even stronger, however model dependent, constraints on the reheat temperature. Combining these arguments yields a constraint on the reheat temperature of typically $T_{RH} < 10^7 - 10^{10} \, \text{GeV}$, depending mainly on the model under consideration and on the value of $m_{3/2}$. The resulting preferred region in ($m_N, \, \tilde{y}_1$)-parameter space is depicted in blue in Fig.~\ref{fig_infl_lepto}.

\subsection{Remarks on Preheating}\label{Remarks on Preheating}
Note that throughout this chapter we have focussed on the evolution of the homogeneous fields $N$ and $H$. It has been pointed out that under certain circumstances this might not be sufficient, since modes with $k \neq 0$ of all fields in the model can be strongly excited at the end of inflation and before the beginning of reheating in a process referred to as preheating. There are two types of preheating worth mentioning in the context of hybrid inflation, namely preheating via parametric resonance~\cite{Kofman:1997yn, BasteroGil:1999fz, GarciaBellido:2000dc, Kofman:1994rk} and tachyonic preheating~\cite{GarciaBellido:2001cb, Felder:2000hj}. In the former case, the coupling of fermions and bosons to the oscillating inflaton field results in oscillating mass terms for these particles. Solving the respective equations of motion (roughly the equation of an harmonic oscillator with an oscillating mass as described, e.g., by the Mathieu equation) can yield explosive particle production. However, in the region of parameter space of interest to us, any heavy particles that are produced by this mechanism will decay back into heavy (s)neutrinos or into radiation. The radiation produced directly or indirectly through this process at the beginning of the reheating phase will however be strongly diluted during the ongoing matter dominated phase and thus be insignificant for the outcome of the reheating phase. Thus in our model, parametric resonance will not affect the results discussed above, mainly due to the structure of the mass spectrum and the very small effective Yukawa coupling $\tilde{y}_1$.

Tachyonic preheating occurs when the squared mass of the $H$ field becomes negative, triggering the waterfall ending inflation. Modes of the $H$ field with $k < |m^{(S)}_H|$ grow exponentially \footnote{It was pointed out in~\cite{McDonald:2001iv} that in some hybrid inflation models a fragmentation of the inflaton condensate can occur, causing the evolution of the universe to be dominated by these 'lumps' instead of by the homogeneous component of the inflaton field. However this 'lump'  formation requires a flatter than $\phi^2$ potential (with $\phi = \{N, \ H\}$), which does not appear in our model as can easily be checked from Eq.~\eqref{global_scalar_pot}.}, causing particle production of bosonic and fermionic fields coupled to the waterfall field~\cite{GarciaBellido:2001cb} and creating an inhomogeneous field $H(x,t)$ which can cause the formation of topological defects when the waterfall occurs~\cite{Felder:2000hj}. It was stated in~\cite{Felder:2000hj} that the production of fermions and bosons coupling to the waterfall field with a coupling strength $g$ is suppressed by $\rho_{B,F}/\rho_V \sim 10^{-3} \, g$ with $\rho_V$ denoting the energy density during inflation. Thus in the parameter region of interest, this is negligible in our model. On the other hand, the production of topological defects could indeed dominate the evolution of the universe in an early stage. However, since we have not observed any topological defects yet, a mechanism to prevent or dilute these objects (e.g.\ a preferred waterfall direction or a slight shift of the potential energy of the discrete vacua) is typically implemented.  We will assume that the higher dimensional operators denoted by dots in Eq.~\eqref{eq_W} provide such a solution so that at some time after the waterfall, the universe is dominated by the lightest right-handed sneutrino. The evolution from this point on is correctly described by the classical theory of reheating, as discussed above. Other possible scenarios in which the evolution of the universe may not be dominated by the homogeneous component of the inflaton field remain to be explored in this context.

\section{Summary and Conclusions: Combining Inflation and Leptogenesis \label{infl_lepto}}

\begin{figure}
\centering
\subfigure[case: $\kappa_{SH} < 1$]{\includegraphics[width=0.49\textwidth]{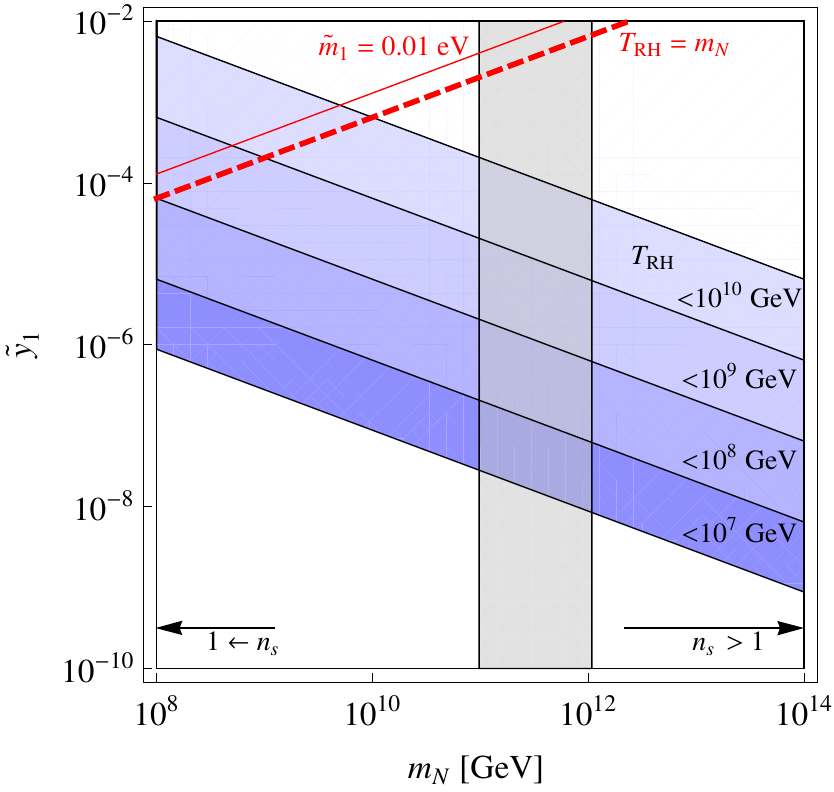}} \hfill
\subfigure[case: $\kappa_{SH} > 1$]{\includegraphics[width=0.49\textwidth]{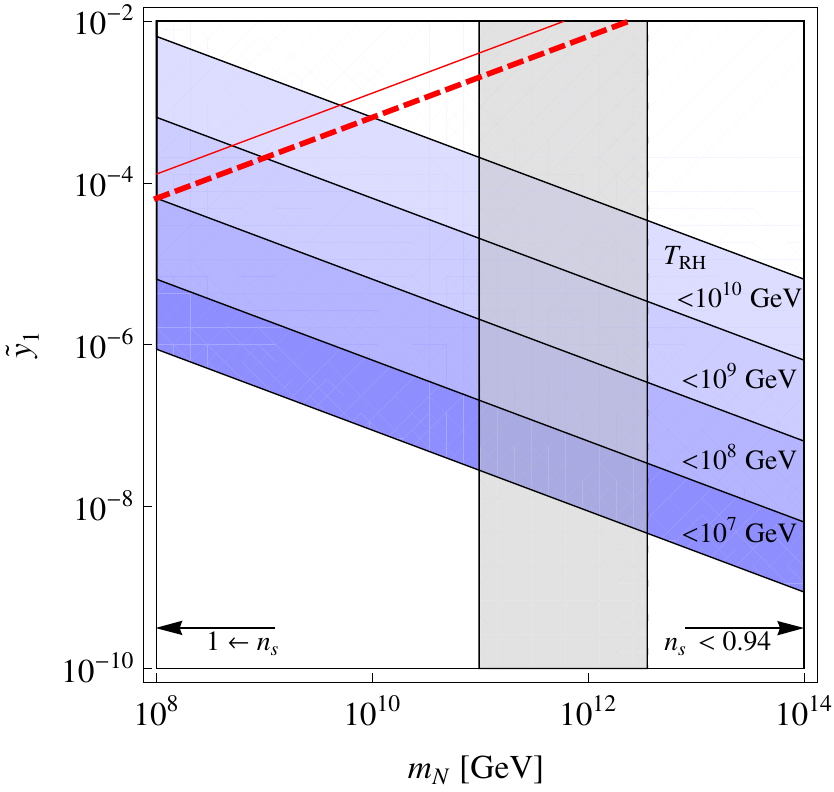}}
\caption{Preferred region (95 $\%$ CL) in ($m_N$, $\tilde{y}_1$)-parameter space from inflation and leptogenesis. The preferred region derived from inflation, in particular from the WMAP constraints on the spectral index $n_s$, is marked in grey. The favored region obtained from reheating and leptogenesis is depicted in blue. The lower bound corresponds to the baryon asymmetry measured by WMAP assuming a maximal CP-violation $\epsilon_{1}$. The gravitino problem imposes a (model dependent) upper bound on the reheat temperature $T_{RH}$. This yields the different shadings, corresponding to a different reheat temperature and correspondingly a different CP-violation. Finally, the red lines depict a constant effective neutrino mass parameter $\widetilde{m}_1$. The upper line corresponds to a light neutrino mass of ${\cal O}(\sqrt{\Delta m^2_{\text{atm, sol}}})$, whereas the dashed line depicts the borderline between predominantly thermal and nonthermal leptogenesis. For these plots we chose $\kappa = 1$. (a) shows the situation for $\kappa_{SH} - 1= -1$. Note that after reaching a minimum at $n_s \simeq 0.98$, the spectral index acquires large values for increasing $m_N$. (b) depicts the situation for $\kappa_{SH} - 1= +1$. In this case, the sign of the SUGRA corrections flips and $n_s$ decreases for large values of $m_N$.}
\label{fig_infl_lepto}
\end{figure}

In sections~\ref{inflation} and~\ref{leptogenesis} we have investigated the conditions under which inflation, with primordial perturbations in accordance with the latest WMAP results, as well as successful leptogenesis can be realized simultaneously in simple models of sneutrino hybrid inflation as outlined in section~\ref{framework}. The combined results are summarized in Fig.~\ref{fig_infl_lepto}.

The dynamics of inflation is governed by the scale $M$ of the phase transition ending hybrid inflation, the mass of the lightest right-handed (s)neutrino $m_N$, the vacuum energy parameter $\kappa$ (= waterfall field self coupling) and the parameter $\kappa_{SH}$ controlling the SUGRA corrections. In principle, terms depending on the neutrino Yukawa coupling matrix could contribute, too. However, in our case the comparatively small first generation Yukawa couplings make these contributions negligible. With $M$ fixed by the amplitude of the scalar CMB fluctuations and $\kappa \sim {\cal O} (1)$, the spectrum of the CMB fluctuations is primarily dependent on the lightest right-handed (s)neutrino mass $m_N$. For large values of $m_N$, SUGRA corrections controlled by $\kappa_{SH}$ become important, with the sign of these contributions depending on the sign of $\kappa_{SH} - 1$. For the spectral index, its running and the tensor-to-scalar ratio the predictions are shown in Fig.~\ref{fig_pred}. Recent WMAP observations constrain the preferred region for the spectral index $n_s$, thus imposing a constraint on the preferred region for $m_N$. For $\kappa = 1$ and $\kappa_{SH} -1 = \pm 1$ this is marked in grey in Fig.~\ref{fig_infl_lepto}.

On the other hand, the decisive quantities of reheating and leptogenesis, namely the reheat temperature $T_{RH}$ and the baryon asymmetry $n_B/n_{\gamma}$ depend on the effective first generation neutrino Yukawa coupling $\tilde{y}_1$, the CP asymmetry $\epsilon_1$ and the mass of the lightest right-handed (s)neutrino $m_N$ 
(see Eqs.~\eqref{Trh} and~\eqref{nB}). The latter parameter is thus the link between inflation and leptogenesis. 
The preferred region of parameter space resulting from bounds on these quantities is marked in blue in Fig.~\ref{fig_infl_lepto}. It is bounded from below by the experimental value of the baryon asymmetry measured by WMAP and by an upper bound on the CP-violation per (s)neutrino decay Eq.~\eqref{epsilon}. From above it is bounded by constraints imposed on the reheat temperature from the gravitino problem. Since these are model dependent, we have plotted the regions satisfying $T_{RH} < 10^{10}, \, 10^9, \, 10^8, \, 10^7$ GeV in different shadings. Note that a higher reheat temperature at a fixed value for $m_N$ automatically corresponds to a smaller value of $\epsilon_1$ in order to match the measured baryon asymmetry. 
The resulting preferred region in parameter space implies an effective first generation Yukawa coupling $\tilde{y}_1 = {\cal O}(10^{-9} - 10^{-4})$. The upper part of this range is of the same order as the first family quark and charged lepton Yukawa couplings, which in the MSSM with moderate $\tan \beta$ are of the order $10^{-4}$ to $10^{-6}$. 

Throughout this paper, we have assumed nonthermal leptogenesis and hierarchical masses of left-handed as well as right-handed neutrinos. Assuming that the light neutrinos obtain masses via a type I seesaw mechanism\footnote{This implies a mass matrix for the left-handed neutrinos $(m_{\nu})_{ij} = -(y_{\nu}^T M^{-1} y_{\nu})_{ij} \langle v \rangle^2/2$.}, both assumptions depend on the value of the effective light neutrino mass parameter (also dubbed washout parameter) $\widetilde{m}_1 \equiv \tilde{y}_1^2 \langle v \rangle^2 / m_N$. More explicitly, one can easily see from Eq.~\eqref{Trh} that $(T_{RH}/m_N)^2 \approx 4.0 \cdot 10^2 \, \widetilde{m}_1/\text{eV}$.
Lines of constant $\widetilde{m}_1$ are marked in red in Fig.~\ref{fig_infl_lepto}, corresponding to a fixed relation between $T_{RH}$ and $m_N$. Simultaneously, they give the order of magnitude for the mass of the left-handed neutrino $m_{\nu_1} \sim \widetilde{m}_1$. In the preferred region of parameter space, we find $\widetilde{m}_1 < 3.4 \cdot 10^{-5}$ eV, thus implying nonthermal leptogenesis with $T_{RH} \ll m_N$ and $m_{\nu_1} \ll \sqrt{\Delta m^2_{\text{atm, sol}}}$.

Finally, we want to comment on possible extensions of this scenario and the significance of cosmological observations in the near future. In Fig.~\ref{fig_infl_lepto} we have set $\kappa = 1$. Allowing for $0.5 < \kappa < 2$ gives qualitatively the same picture (see section~\ref{inflation}, in particular Fig.~\ref{fig_pred}) with a somewhat shifted grey region. 
For example, for $\kappa = 2$ the grey region is extended to the left to $m_N = 2 \cdot 10^{10}$ GeV whereas for $\kappa = 0.5$ it is extended to the right to $m_N = 7 \cdot 10^{12}$ GeV for $\kappa_{SH} -1= +1$.\footnote{For $\kappa_{SH} = -1$ the respective region is extended to $m_N = 2 \cdot 10^{12}$ GeV.} Another interesting possibility would arise if the experimentally preferred region for the spectral index was raised, favoring a spectral index closer to 1. This would lower the preferred range for the lightest (s)neutrino mass $m_N$ significantly and thus open up the region of thermal leptogenesis and allow for  $m_{\nu_1} \sim {\cal O}(\sqrt{\Delta m^2_{\text{atm, sol}}})$. The forthcoming results of the Planck satellite will make the requirements for $m_N$ more accurate.  

In summary, we have pointed out that successful sneutrino hybrid inflation and leptogenesis can be achieved in this framework, and that combining both imposes requirements on the parameters of the underlying particle physics model. We obtain a mass for the lightest right-handed (s)neutrino of $m_N = {\cal O}(10^{10} - 10^{13})$ GeV, an effective first generation neutrino Yukawa coupling $\tilde{y}_1 = {\cal O}(10^{-9} - 10^{-4})$ and a very light left-handed neutrino with $m_{\nu_1} < {\cal O} (10^{-4})$ eV. Furthermore, we find that leptogenesis occurs via nonthermal leptogenesis (with $T_{RH}/m_N < 0.1$ for $\kappa = 1$). Concerning the spectrum of the CMB fluctuations, we predict a running of the spectral index of $-0.0004 < \alpha_s < 0.0002$ and a tensor-to-scalar ratio of $r \lesssim 0.015$. Our results provide a guideline for the construction of explicit particle physics models incorporating sneutrino hybrid inflation and subsequent baryogenesis via nonthermal leptogenesis.

\section*{Acknowledgments}
We would like to thank Alejandro Ibarra and Koushik Dutta for discussions. We acknowledge partial
support by the DFG cluster of excellence ``Origin and Structure of the Universe''.


\begin{thebibliography}{999}


\bibitem{Guth:1980zm}
A.~H. Guth,
Phys. Rev. \textbf{D23} (1981), 347--356;
%
A.~D. Linde,
Phys. Lett. \textbf{B108} (1982), 389--393;
%
A.~Albrecht and P.~J. Steinhardt,
Phys. Rev. Lett. \textbf{48}
 (1982), 1220--1223;
%
for a review containing an extensive list of references,
see e.g.: D.~H.~Lyth and A.~Riotto,
Phys.\ Rept.\  {\bf 314} (1999) 1.  

\bibitem{Liddle:2000cg}
For textbook reviews on inflation see:
A.~R.~Liddle and D.~H.~Lyth, 
``Cosmological inflation and large-scale structure,''
{\it  Cambridge, UK: Univ. Pr. (2000) 400 p};
%
A.~D.~Linde, 
``Particle Physics and Inflationary Cosmology,''
[arXiv:hep-th/0503203];
%
V.~Mukhanov, ``Physical Foundations of Cosmology,''
{\it  Cambridge, UK: Univ. Pr. (2005) 421 p}

\bibitem{Bailin:2004zd} 
  D.~Bailin and A.~Love,
{\it  Bristol, UK: IOP (2004) 313 p}.


\bibitem{Baumann:2009ds}
  D.~Baumann,
  arXiv:0907.5424 [hep-th].
  
\bibitem{Mazumdar:2010sa}
  A.~Mazumdar and J.~Rocher,
  arXiv:1001.0993 [hep-ph].


\bibitem{Fukugita:1986hr}
  M.~Fukugita and T.~Yanagida,
  Phys.\ Lett.\  B {\bf 174}, 45 (1986).

\bibitem{seesaw}
P.~Minkowski,
Phys.\ Lett.\ B {\bf 67} (1977) 421;
M. Gell-Mann, P. Ramond and R. Slansky in Sanibel Talk,
CALT-68-709, Feb 1979, and in {\it Supergravity} (North Holland,
Amsterdam 1979);
T. Yanagida in {\it Proc. of the Workshop on Unified Theory and
Baryon Number of the Universe}, KEK, Japan, 1979;
S.L.Glashow, Cargese Lectures (1979);
R.~N.~Mohapatra and G.~Senjanovic,
Phys.\ Rev.\ Lett.\  {\bf 44} (1980) 912;
J.~Schechter and J.~W.~Valle,
Phys.\ Rev.\ D {\bf 25} (1982) 774.

\bibitem{Chen:2007fv}
  M.~C.~Chen,
  arXiv:hep-ph/0703087.

\bibitem{Davidson:2008bu}
  S.~Davidson, E.~Nardi and Y.~Nir,
  Phys.\ Rept.\  {\bf 466}, 105 (2008)
  [arXiv:0802.2962 [hep-ph]].

\bibitem{Murayama:1992ua}
 H.~Murayama, H.~Suzuki, T.~Yanagida and J.~Yokoyama,
 Phys.\ Rev.\ Lett.\ {\bf 70} (1993) 1912.

\bibitem{Ellis:2003sq}
  J.~R.~Ellis, M.~Raidal and T.~Yanagida,
  Phys.\ Lett.\  B {\bf 581}, 9 (2004)
  [arXiv:hep-ph/0303242].

\bibitem{Antusch:2004hd}
  S.~Antusch, M.~Bastero-Gil, S.~F.~King and Q.~Shafi,
  Phys.\ Rev.\  D {\bf 71}, 083519 (2005)
  [arXiv:hep-ph/0411298].


\bibitem{Linde:1990gz}
  A.~D.~Linde,
  Phys.\ Lett.\  B {\bf 249}, 18 (1990).

\bibitem{Linde:1991km}
  A.~D.~Linde,
  Phys.\ Lett.\  B {\bf 259}, 38 (1991).

\bibitem{Copeland:1994vg}
  E.~J.~Copeland, A.~R.~Liddle, D.~H.~Lyth, E.~D.~Stewart and D.~Wands,
  Phys.\ Rev.\  D {\bf 49}, 6410 (1994)
  [arXiv:astro-ph/9401011].

\bibitem{Linde:1997sj}
  A.~D.~Linde and A.~Riotto,
  Phys.\ Rev.\  D {\bf 56}, 1841 (1997)
  [arXiv:hep-ph/9703209].

\bibitem{Antusch:2009vg}
  S.~Antusch, K.~Dutta and P.~M.~Kostka,
  AIP Conf.\ Proc.\  {\bf 1200}, 1007 (2010)
  [arXiv:0908.1694 [hep-ph]].
  
\bibitem{Kawasaki:2000yn}
 M.~Kawasaki, M.~Yamaguchi and T.~Yanagida,
 Phys.\ Rev.\ Lett.\  {\bf 85}, 3572 (2000)
 [arXiv:hep-ph/0004243].

\bibitem{Yamaguchi:2000vm}
 M.~Yamaguchi and J.~Yokoyama,
 Phys.\ Rev.\  D {\bf 63}, 043506 (2001)
 [arXiv:hep-ph/0007021].

\bibitem{Kawasaki:2000ws}
 M.~Kawasaki, M.~Yamaguchi and T.~Yanagida,
 Phys.\ Rev.\  D {\bf 63}, 103514 (2001)
 [arXiv:hep-ph/0011104].

\bibitem{Antusch:2009ef}
  S.~Antusch, K.~Dutta and P.~M.~Kostka,
  Phys.\ Lett.\  B {\bf 677}, 221 (2009)
  [arXiv:0902.2934 [hep-ph]].

\bibitem{Antusch:2008pn}
  S.~Antusch, M.~Bastero-Gil, K.~Dutta, S.~F.~King and P.~M.~Kostka,
  JCAP {\bf 0901}, 040 (2009)
  [arXiv:0808.2425 [hep-ph]].

\bibitem{Antusch:2010va}
  S.~Antusch, M.~Bastero-Gil, J.~P.~Baumann, K.~Dutta, S.~F.~King and P.~M.~Kostka,
  arXiv:1003.3233 [hep-ph].

\bibitem{Antusch:2008gw}
S.~Antusch, S.~F.~King, M.~Malinsky, L.~Velasco-Sevilla and
I.~Zavala,
Phys.\ Lett.\ B {\bf 666}, 176 (2008)
[arXiv:0805.0325 [hep-ph]].

\bibitem{Percival}  W.~J.~Percival {\it et al.},
  Mon.\ Not.\ Roy.\ Astron.\ Soc.\  {\bf 401}, 2148 (2010)
  [arXiv:0907.1660 [astro-ph.CO]].

\bibitem{Riess}   A.~G.~Riess {\it et al.},
  Astrophys.\ J.\  {\bf 699}, 539 (2009)
  [arXiv:0905.0695 [astro-ph.CO]].

\bibitem{wmap} WMAP cosmological parameters. Model: lcdm+sz+lens. Data: wmap7+bao+h0\\
http://lambda.gsfc.nasa.gov/product/map/current/params/ lcdm$\_$sz$\_$lens$\_$wmap7$\_$bao$\_$h0.cfm

\bibitem{coleman_weinberg_potential}

 E.~J.~Weinberg,
  arXiv:hep-th/0507214.

  G.~Gamberini, G.~Ridolfi and F.~Zwirner,
  Nucl.\ Phys.\  B {\bf 331} (1990) 331.

 S.~Weinberg,
  Phys.\ Rev.\  D {\bf 7}, 2887 (1973).


\bibitem{Dine:1995uk}
  M.~Dine, L.~Randall and S.~D.~Thomas,
  Phys.\ Rev.\ Lett.\  {\bf 75}, 398 (1995)
  [arXiv:hep-ph/9503303].


\bibitem{Kofman:1997yn}
  L.~Kofman, A.~D.~Linde and A.~A.~Starobinsky,
  Phys.\ Rev.\  D {\bf 56}, 3258 (1997)
  [arXiv:hep-ph/9704452].

\bibitem{Allahverdi:2005mz}
  R.~Allahverdi and A.~Mazumdar,
  JCAP {\bf 0610}, 008 (2006)
  [arXiv:hep-ph/0512227].

\bibitem{Allahverdi:2007zz}
  R.~Allahverdi and A.~Mazumdar,
  Phys.\ Rev.\  D {\bf 76}, 103526 (2007)
  [arXiv:hep-ph/0603244].



\bibitem{Covi:1996wh}
L.~Covi, E.~Roulet and F.~Vissani,
 Phys.\ Lett.\ B {\bf 384} (1996) 169
 [arXiv:hep-ph/9605319].

\bibitem{Davidson:2002qv}
  S.~Davidson and A.~Ibarra,
  Phys.\ Lett.\  B {\bf 535}, 25 (2002)
  [arXiv:hep-ph/0202239].

\bibitem{Hamaguchi:2001gw}
  K.~Hamaguchi, H.~Murayama and T.~Yanagida,
  Phys.\ Rev.\  D {\bf 65}, 043512 (2002)
  [arXiv:hep-ph/0109030].


\bibitem{Kolb:1990vq}
  E.~W.~Kolb and M.~S.~Turner,
  Front.\ Phys.\  {\bf 69}, 1 (1990).

\bibitem{Komatsu:2010fb}
  E.~Komatsu {\it et al.},
  arXiv:1001.4538 [astro-ph.CO].

\bibitem{Khlopov:1984pf}
  M.~Y.~Khlopov and A.~D.~Linde,
  Phys.\ Lett.\  B {\bf 138}, 265 (1984).

\bibitem{Ellis:1984eq}
  J.~R.~Ellis, J.~E.~Kim and D.~V.~Nanopoulos,
  Phys.\ Lett.\  B {\bf 145}, 181 (1984).

\bibitem{Ellis:1984er}
  J.~R.~Ellis, D.~V.~Nanopoulos and S.~Sarkar,
  Nucl.\ Phys.\  B {\bf 259}, 175 (1985).

\bibitem{Moroi:1993mb}
  T.~Moroi, H.~Murayama and M.~Yamaguchi,
  Phys.\ Lett.\  B {\bf 303}, 289 (1993).

\bibitem{Kawasaki:2004yh}
  M.~Kawasaki, K.~Kohri and T.~Moroi,
  Phys.\ Lett.\  B {\bf 625}, 7 (2005)
  [arXiv:astro-ph/0402490].

\bibitem{BasteroGil:1999fz}
  M.~Bastero-Gil, S.~F.~King and J.~Sanderson,
  Phys.\ Rev.\  D {\bf 60}, 103517 (1999)
  [arXiv:hep-ph/9904315].

\bibitem{GarciaBellido:2000dc}
  J.~Garcia-Bellido, S.~Mollerach and E.~Roulet,
  JHEP {\bf 0002}, 034 (2000)
  [arXiv:hep-ph/0002076].


\bibitem{Kofman:1994rk}
  L.~Kofman, A.~D.~Linde and A.~A.~Starobinsky,
  Phys.\ Rev.\ Lett.\  {\bf 73}, 3195 (1994)
  [arXiv:hep-th/9405187].


\bibitem{GarciaBellido:2001cb}
  J.~Garcia-Bellido and E.~Ruiz Morales,
  Phys.\ Lett.\  B {\bf 536}, 193 (2002)
  [arXiv:hep-ph/0109230].

\bibitem{Felder:2000hj}
  G.~N.~Felder, J.~Garcia-Bellido, P.~B.~Greene, L.~Kofman, A.~D.~Linde and I.~Tkachev,
  Phys.\ Rev.\ Lett.\  {\bf 87}, 011601 (2001)
  [arXiv:hep-ph/0012142].
  
\bibitem{McDonald:2001iv}
  J.~McDonald,
  Phys.\ Rev.\  D {\bf 66}, 043525 (2002)
  [arXiv:hep-ph/0105235].

 \end{thebibliography}
\end{document}